# Band-9 ALMA observations of the [NII] 122 µm line and FIR continuum in two high-z galaxies.

Carl Ferkinhoff[1,2], Drew Brisbin[3,2], Thomas Nikola[3], Gordon J. Stacey[2], Kartik Sheth[4], Steve Hailey-Dunsheath[5], Edith Falgarone[6]

[1] Max-Planck-Institut für Astronomie, Königstuhl 17, D-69117 Heidelberg, Germany; ferkinhoff@mpia.de
[2] Department of Astronomy, Cornell University, Ithaca, NY 14853, USA
[3] Center for Radiophysics and Space Research, Cornell University, Ithaca, NY 14853, USA
[4] National Radio Astronomy Observatory, Charlottesville, VA 22903, USA
[5] California Institute of Technology, Mail Code 301-17, 1200 E. California Blvd., Pasadena, CA 91125, USA
[6] LERMA, CNRS, Observatoire de Paris and ENS, France

**Abstract:**

We present Atacama Large Millimeter Array (ALMA) observations of two high-redshift systems (SMMJ02399-0136 and the Cloverleaf QSO) in their rest-frame 122 µm continuum ($\nu_{sky}$~650 GHz, $\lambda_{sky}$~450 µm) and [NII] 122 µm line emission. The continuum observations with a synthesized beam of ~0.25" resolve both sources and recover the expected flux. The Cloverleaf is resolved into a partial Einstein ring, while the SMMJ02399-0136 is unambiguously separated into two components; an AGN associated point source and an extend region at the location of a previously identified dusty starburst. We detect the [NII] line in both systems, though significantly weaker than our previous detections made with the 1st generation z(Redshift) and Early Universe Spectrometer. We show that this discrepancy is mostly explained if the line flux is resolved out due to significantly more extended emission and longer ALMA baselines than expected. Based on the ALMA observations we determine that ≥75% of the total [NII] line flux in each source is produced via star formation. We use the [NII] line flux that is recovered by ALMA to constrain the N/H abundance, ionized gas mass, hydrogen ionizing photon rate, and star formation rate. In SMMJ02399-0136 we discover it contains a significant amount (~1000 $M_\odot$ yr$^{-1}$) of unobscured star formation in addition to its dusty starburst and argue that SMMJ02399-0136 may be similar to the Antennae Galaxies (Arp 244) locally. In total these observations provide a new look at two well-studied systems while demonstrating the power and challenges of Band-9 ALMA observations of high-z systems.

Keywords: galaxies: active—galaxies: high-redshift—galaxies: individual (SMM J02399-0136; H11413+117)—galaxies: ISM—submillimeter: galaxies

1.  **Introduction**

The fine-structure emission lines of oxygen, nitrogen, carbon and their various ions that are emitted in the far-infrared (FIR) regime are important probes of a galaxy's interstellar medium (ISM) and radiation fields (Carilli & Walter 2012). The lines are insensitive to gas temperature and often optically thin. They are also minimally affected by extinction due to dust, especially when compared to the extinction at optical wavelengths, making them especially useful for studies of dusty systems like luminous (LIRG; $L_{IR} > 10^{11}$ $L_\odot$) and ultra-luminous infrared galaxies (ULIRG; $L_{IR} > 10^{12}$ $L_\odot$). The far-infrared dust continuum, which probes a galaxy's dust obscured star formation, is vital for understanding the rate of star formation in these same systems (Casey, Narayanan & Cooray 2014). It is also often detected along with the FIR lines, which is not surprising given that the peak in the dust SED occurs at or near where the lines are emitted. For sources at redshifts ≳ 1, the FIR lines and continuum fall into the submillimeter regime making them observable with ground based facilities through the submillimeter telluric windows. These qualities make the FIR lines and continuum ideal for studies of early galaxies.

Over the past decade the first studies of the FIR fine-structure lines from high-z galaxies were performed (e.g. Maiolino et al. 2005, 2009; Hailey-Dunsheath et al. 2010; Stacey et al 2010; Ferkinhoff et al. 2010, 2011, 2014; Decarli et al. 2012, 2014; Brisbin et al. 2015). These important and pioneering works demonstrate the potential of these emission lines as probes of the physical conditions of high-z galaxies. The Atacama Large Millimeter Array (ALMA) is poised to revolutionize these studies. ALMA represents an order-of-magnitude advance in terms of both sensitivity and spatial resolution for observations in the submillimeter providing the potential to bring studies of FIR fine-structure lines to many hundreds, if not thousands, of galaxies in the early universe. Here we present observations of the [NII] 122 μm line and rest frame 122 μm continuum emission of two high-z galaxies that demonstrate ALMA's capabilities for FIR studies of systems in the early universe. Considering that these observations were made with only ~20 antennas and that ALMA is now "routinely" detecting the [CII] 158 μm line from high-z galaxies, the full ALMA array of fifty antenna will enable enormous advances in this field.

Using Band-9 of ALMA ($\nu_{sky}$ ~ 650 GHz, $\lambda_{sky}$ ~ 450 μm) in Cycle-0 we mapped the 122 μm rest-frame continuum of SMMJ02399-0136 (hereafter SMMJ02399; Figure 1) and the Cloverleaf QSO (H1413+117; Figure 2) at z~ 2.8 and 2.5 respectively with a resolution of ~0.25". We also detected the 122 μm FIR fine-structure line of singly ionized nitrogen ([NII] 122 μm) in both sources, confirming our previous detections made with the 1$^{st}$ generation z(Redshift) and Early Universe Spectrometer (ZEUS-1, Hailey-Dunsheath 2009), albeit at significantly lower flux.

In this paper we first describe the utility of the [NII] 122 μm line and summarize the results of previous studies of our sources. In section 2 we describe our new ALMA observations, their reduction and calibration. Section 3 describes the production of the [NII] 122 μm line and continuum maps, and the extracted spectra. In Section 4 we investigate the nature of the [NII] line. Section 5 details our analysis of the continuum and spectral data while in section 6 we discuss our results and their impact on our understanding of the sources and high-z sources more broadly. Lastly we summarize our key findings and their implications for future studies with ALMA in Section 7. Throughout this paper, we adopt the cosmological parameters of $\Omega_\Lambda$ =0.73, $\Omega_m$ =0.27 and $H_0$ = 71 km s$^{-1}$ Mpc$^{-1}$ (Komatsu et al. 2011).

*1.1    The utility of the 122 μm fine-structure emission of ionized nitrogen.*

The [NII] 122 μm line arises in the electronic ground-state of singly ionized nitrogen due to transitions between the $^3P_2$ and $^3P_1$ fine-structure levels. As ≥14.5 eV photons are required to produce singly ionized nitrogen, the [NII] 122 μm line traces only the ionized gas of either stellar HII regions formed by O & B stars or the narrow-line-region (NLR) produced by AGN (Active Galactic Nucleus). The upper level has an equivalent temperature above ground of 188 K and the transition has a critical density of 310 cm$^{-3}$. Due to the low equivalent temperature required to populate the emitting level, the line is insensitive to the temperatures encountered in the HII regions. Furthermore the modestly high critical density means the line is often emitted in the low-density regime so its line flux constrains the minimum number of hydrogen-ionizing photons necessary to maintain the ionization equilibrium in the HII region. On the other hand, if one assumes the line emission arises in the high-density regime—where the gas density is larger than the emitting levels critical density—then the line flux constrains the minimum gas mass necessary to produce the observed emission. Both constraints—the ionizing photon rate and ionized gas mass—are also affected by assumptions on the nitrogen-to-hydrogen gas-phase abundance and ionization state of the emitting gas. The latter determines the fraction of nitrogen that is singly ionized and can have large effects on the estimates of the ionized gas-mass and photon-rate.

Comparing the [NII] line strength to lines emitted by higher ionization species (e.g. the [OIII] 88 μm line) is one way to constrain the ionization state of the ionized region since the comparison determines the hardness of the ionizing radiation field. In the NLRs, where the UV hardness varies little for different power-law indices of the ionizing source, the [OIII]/[NII] line ratio determines the ionization parameter—the ratio of the number of ionizing photons to the number of hydrogen nuclei. For stellar HII regions the line ratio gives a direct measure of the effective stellar temperature of the stars responsible for the ionization. This constrains the spectral type of the most massive star still on the main sequence and subsequently the age of the starburst. Furthermore, we can determine the number of the HII regions and subsequently the number of the most-massive stars by comparing the expected line luminosity for a given HII region model to the total observed line luminosity.

### *1.2 Sources: SMMJ02399 and the Cloverleaf*

The sources in this study, SMMJ02399 at z = 2.8076 and the Cloverleaf QSO at z=2.5579, are both lensed broad-absorption-line (BAL) quasars with multiple components. We summarize the source properties available in the literature in Table 1. The Cloverleaf's components (labeled A, B, C and D; see figure 4) are multiple images of a single background AGN that are produced via strong gravitational lensing (magnification factor μ ~ 11; Venturini et al. 2003) by two foreground lensing galaxies. It is the first confirmed example of a multiply lensed system (Hazard et al. 1984, Magain et al .1988). Subsequent studies showed the Cloverleaf to host a significant molecular gas reservoir and starburst in addition to its AGN (Weiss et al. 2003; Barvainis et al. 1997, Bradford et al. 2009, Henkel et al. 2010). To date, the brightness of the quasar images at visible wavelengths and the low resolution of millimeter observations have been inadequate to resolve any extended emission, i.e. an Einstein ring. Detailed gravitational lens modeling of the CO (7-6) emission suggests that the Cloverleaf contains a molecular gas disk with radius ~650 pc (Venturini & Solomon 2003).

SMMJ02399, the first spectroscopically confirmed submillimeter galaxy (Ivison et al. 1998), is only weakly lensed (μ ~ 2.38; Ivison et al. 2010) by the foreground cluster Abell 370. In total four photometric components in SMMJ02399 have been identified. Unlike the Cloverleaf these are not multiple lensed images, but represent 2 or more individual galaxies that are likely closely associated (within ~10 kpc of each other in the source-plane) and undergoing a merger (Ivison et al. 2010). SMMJ02399 component L1 hosts the quasar while L2SW is the site of a dusty starburst and coincident with the peak on molecular gas emission (Ivison et al. 2010; see Figure 3). The L2 and L1N components contain significant stellar populations (Aguirre et al. 2013) but little dust. Spectroscopic observations of CO gas suggest that there is a velocity gradient across SMMJ02399, with L2 and L2SW at approximately -500 km s$^{-1}$ and L1 near +250 km s$^{-1}$ relative to the overall systemic redshift (Genzel et al. 2003).

Owing to both the Cloverleaf's and SMMJ02399's large apparent FIR luminosities, molecular gas reservoirs and star formation rates we identified them as ideal targets for observations of their FIR fine-structure emission lines with our spectrometer ZEUS-1 on the Caltech Submillimeter Observatory. Indeed, we detected the [NII] 122 μm line in both SMMJ02399 and the Cloverleaf (2.47 × 10$^{-18}$ and 2.42 × 10$^{-18}$ W m$^{-2}$ or 130 and 105 Jy km s$^{-1}$ respectively; Ferkinhoff et al. 2011) as well as the [OIII] 88 μm line in SMMJ02399 (6.04 × 10$^{-18}$ W m$^{-2}$ or 200 Jy km s$^{-1}$; Ferkinhoff et al. 2010).[1,2]

In Ferkinhoff et al. (2011) we use the [NII] line flux to show that both SMMJ02399 and the Cloverleaf have a significant quantity of ionized gas (~3.3 × 10$^{10}$ M$_\odot$ and ~2.1 × 10$^9$ M$_\odot$), equivalent to ~ 14% and 7% of their molecular gas respectively. In the case of SMMJ02399, for which we also detected the [OIII] line, we determine the [NII] emission is produced in (1 – 3) × 10$^8$ HII regions that are ionized by O9.5 stars—constraining the starburst to be no more than 10 million years old. Unfortunately, the observed [NII] and [OIII] line strengths in SMMJ02399 are also consistent with being produced in a NLR or even a composite of NLR and star formation. This is also true for the [NII] emission in the Cloverleaf. Since the [NII] line is the only fine-structure line observed in the Cloverleaf, the sources of the emission—starburst or AGN—is not constrained. However, given the intense starburst known to reside in the Cloverleaf, we assumed a star forming paradigm and compare the [NII] line along with other star formation tracers (e.g. PAH emission, Hα, molecular gas mass, etc.) in the Cloverleaf to those found in nearby starbursting system M82. We find that both systems have similar ratios, albeit the Cloverleaf is significantly more luminous. It can be described as a superposition of two-hundred M82 like starbursts. If this model is correct, then the Cloverleaf's starburst has a similar age and number of HII regions as SMMJ02399.

---

[1] Unless noted otherwise all direct observables (e.g. flux density, line flux, etc.) are the apparent values, uncorrected for lensing, while all derived values (e.g. masses and luminosities) are the intrinsic source values corrected by the lensing factors listed in table 1.

[2] The [NII] line fluxes from ZEUS-1 used here are 20% lower than the values reported in Ferkinhoff et al. 2011 due to an improved flux calibration.

Given that both systems are known to contain an AGN and starburst and that our FIR fine-structure line observations are consistent with both AGN and starburst origins, we cannot conclusively determine the nature of their [NII] emission. Additional spectral probes from even higher ionization states (e.g. [OIV]) would help to clearly identify the source of the emission as has been demonstrated with Herschel observations of nearby galaxies (Spinoglio et al. 2015). Alternatively, since the NLR region should be confined to within a few hundred parsecs of the AGN, while the star formation is likely to extend over kiloparsec scales, high spatial-resolution mapping can also elucidate the nature of the [NII] emission. With this in mind we proposed to obtain high resolution imaging of the [NII] 122 μm line using Band-9 of ALMA during the cycle-0 early science operations.

*Table 1: Source Properties*

| Source Properties | Units | SMMJ02399 | Ref. | The Cloverleaf | Ref |
|---|---|---|---|---|---|
| (1) | (2) | (3) | (4) | (5) | (6) |
| R.A. | … | 02h39m51.9s | … | 14h15m46.3s | … |
| Dec. | … | −01°35'59" | … | 11°29'44" | … |
| z | … | 2.8076 | Genzel et al. 2003 | 2.5579 | Barvainis et al. 1997 |
| $D_L$ | Gpc | 23.8 | … | 21.3 | … |
| μ | … | 2.38 | Ivison et al. 2010 | 11 | Venturini & Solomon 2003 |
| F([NII]) – ZEUS[a] | $10^{-18}$ W m$^{-2}$ (Jy km /s) | 2.77 ± 0.43 (130 ± 20) | Ferkinhoff et al. 2011 | 2.42 ± 0.40 (105 ± 17) | Ferkinhoff et al. 2011 |
| F([OIII]) - ZEUS | $10^{-18}$ W m$^{-2}$ (Jy km /s) | 6.04 ± 1.46 (200 ± 50) | Ferkinhoff et al. 2010 | … | … |
| $L_{FIR}$[b] | $10^{12} L_\odot$ | 4.0 | Thomson et al. 2012 | 5.58 | Bradford et al. 2009, Lutz et al. 2007 |
| $M_{H2}$ | $10^{10} M_\odot$ | 4.1 ± 1.4 | Ivison et al. 2010 | 2.6 ± 2.4 | Bradford et al. 2009 |
| $M_{dust}$ | $10^8 M_\odot$ | 7-5 | Ivison et al. 1998 | 0.61 (±12%) and 0.0035 (±60%)[c] | Weiß et al. 2003 |
| $T_{dust}$ | K | 40-50 | Ivison et al. 1998 | 50(±2) and 115 (±10)[c] | Weiß et al. 2003 |
| $S_v$(450 μm) | mJy | 69 ± 15 | Ivison et al. 1998 | 210 ± 20 | Weiß et al. 2003 |

Notes: [a] The [NII] line fluxes here are 20% lower than the values reported in Ferkinhoff et al. 2011 due to an improved calibration.
[b] Thomson et al. 2012 determine $L_{IR}$(8 – 1000 μm), the total IR luminosity. $L_{FIR}$ (42.5 – 122.5 μm) ~ 0.5 × $L_{IR}$(8 – 1000 μm)
[c] Weiß et al. (2003) perform a two temperature SED fit to the available Cloverleaf photometry and determine a dust mass and temperature for a cold and warm component respectively.

## 2. Observations and Calibrations

The North American ALMA Science Center (NAASC) provided calibrated data products, raw data, and reduction scripts to reproduce the calibrated data. We reprocessed the raw data, using the scripts provided by the NAASC as a guide. Our final data products and calibration process was largely unchanged from the scripts provided except for a few minor, yet important, differences we describe below.

### 2.1 SMMJ02399

Observations of SMMJ02399 were performed on 28 August 2012. The total integration time was 59.7 minutes using ~27 antennas with baselines between ~40 and 400 meters. 33 minutes were spent on source in good Band-9 observing conditions of 0.42 mm line-of-sight perceptible water vapor (PWV). The remaining time was spent for calibration and pointing. The receivers were tuned and spectral windows placed to provide complete spectral coverage between 643.0 and 650.5 GHz so that the [NII] line would fall near the middle of the frequency range.

Five antennas were flagged and removed from the data before calibration because of high receiver system-temperatures, bad pointing, or in one case, poor antenna position. Flux calibration was done by observing Neptune, while PKS J0538-4405 was used for bandpass calibration and initial phase calibration of the weak phase calibrator ($S_\nu$ ~ 0.29 Jy) PKS J0215+015 so that a phase solution over time could be obtained. We verified the calibration by imaging the bandpass-calibrator, primary phase-calibrator, and the secondary phase-calibrator and confirmed that the observed fluxes and astrometry matched expected values. The final calibrated data and calibration process was identical to that provided by the NAASC, except that we exclude the last scan (~6 minutes of on-source integration time) from the final data cube. The phase noise during this time, on both the phase-calibrators and SMMJ02399, was ~2 times higher than the rest of the observations. Imaging of the data showed that the final image RMS was the same whether the final scan was included or not, yet was achieved in fewer cycles of the cleaning process when the scan was excluded. As such we excluded it from our final data cube.

### 2.2 The Cloverleaf

Observations of the Cloverleaf were executed over three epochs in 2012 on June 5, December 12 and the night of July 16 totaling ~204 minutes of total integration time, 115 of them were spent on source. The June 5 and July 16 observation were performed with 21 antennas in 0.54 and 0.57 mm of line-of-sight PWV respectively. The December 12 observation had 22 antennas in the array and excellent observing conditions with 0.35 mm of PWV. The baselines ranged from ~ 40 meters to ~500 meters. The receivers were tuned and spectral windows placed to provide complete spectral coverage between 688.5 and 699.0 GHz so that the [NII] would fall near the middle of the frequency range.

For each epoch one, three and one antenna were flagged respectively due to either high receiver system-temperatures or bad pointing. As with SMMJ02399 we used the observations of the bandpass calibrator to provide an initial phase calibration for the weak phase calibrator J1413+133 for which we averaged all the spectral windows in order to have enough signal to noise for a successful phase solution over time. Flux calibration was done with observations of Titan, though due to prominent lines from Titan's atmosphere only 2/3 of one spectral window were used to successfully flux calibrate the data set. Flux and astrometry checks on 3C279 (a bright phase calibrator that was ~26 degrees from the Cloverleaf, too far to provide a phase solution) and J1413+133 verified a successful calibration. Furthermore, they were consistent across the three epochs so that we kept the individual calibrations in the final combined data set used for imaging. In producing our final data cube we chose to flag fewer of the spectral channels in the source observations compared to the product from the NAASC. This was done to ensure we had a continuous spectrum across the four spectral windows of the observations. We also chose not to flag the channels containing three weak telluric features. This was important as one feature lies very close to the expected position of the [NII] line and flagging those spectral channels prior to imaging makes identifying the [NII] line difficult. Furthermore, assuming a proper calibration, the presence of the telluric lines should only result in increased noise in those respective channels, which we must remain aware of during imaging and interpretation of our data.

## 3.  Imaging

### 3.1  *Continuum Imaging*

For both sources we imaged the continuum over the entire ~7.5 GHz bandwidth of the four spectral windows. To produce the final continuum maps (Figures 1 and 2) we iteratively cleaned the dirty images in CASA 4.2 down to peak residuals of ~1 mJy using a "natural" weighting scheme of the visibilities. To optimize our signal-to-noise ratio (SNR) and spatial resolution in the maps while assessing if we had resolved-out any flux we repeated the imaging process using several different weighting schemes. Figures 3 and 4 show continuum maps produced with uniform, natural and robust (R = 0.0) weighting schemes as well as an outer UV taper, which produces ~0.5" synthesized beam in the cleaned image. In Table 2 we list the properties of each map, including peak intensities and source integrated fluxes, over SMMJ02399 and the Cloverleaf as well as for their individual components indicated in Figures 3 and 4. Hereafter we will use the component names shown in Figure 3 and 4 when referring to the sources and their respective components.

In the case of SMMJ02399 both the natural weighted (~0.25") and UV tapered (~0.5") maps have approximately the same 1-sigma RMS level of ~1 mJy beam$^{-1}$ as well as the same source-integrated continuum flux-density over the source. Furthermore, the integrated flux-density (~69 mJy) agrees with previous single dish observations at 450 μm (~122 μm rest-frame) noted in Table 1. These factors are important as they suggest that even with a 0.25" beam we are fully recovering the entire continuum flux-density of the source. This is similarly the case for the Cloverleaf where the naturally weighted and UV tapered maps have source-integrated flux-densities that agree with previous single dish observations. However significant emission is still evident in the uniformly weighted map. Most significantly, peak emission in the Cloverleaf changes from component B to component C as the weighting changes from the natural to the uniform weighting. This suggests that the uniform weighted map, with its smaller beam and higher noise level, is probing point-source like emission possibly associated with the AGN, while the natural weighted maps includes an extended component possibly tracing the host galaxy and its star formation. Given, that the natural weighted maps in both sources show the lowest 1-sigma RMS level, we will use the continuum flux-density extracted from those maps from here onwards, unless otherwise noted.

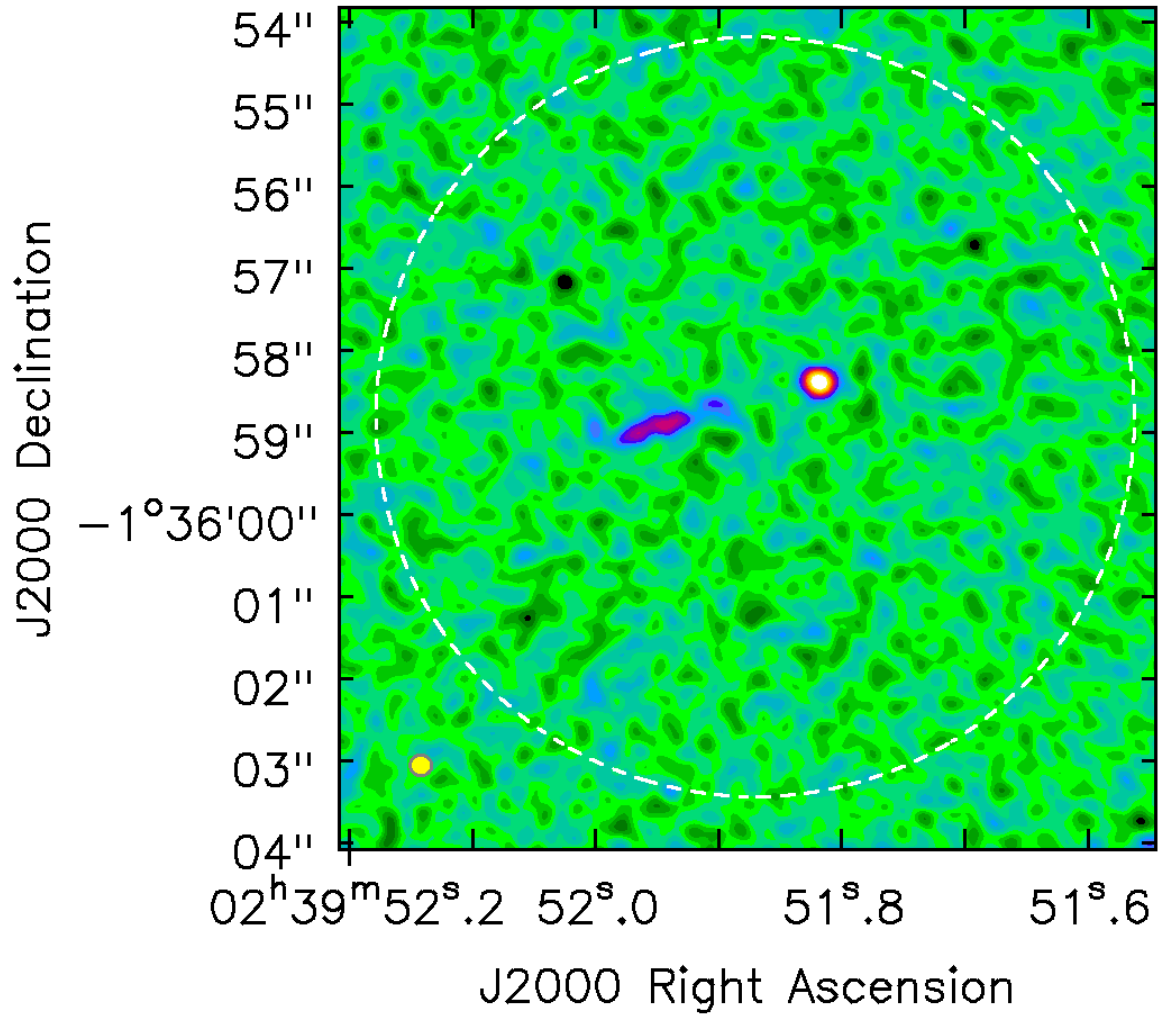

Figure 1: 122 μm rest frame continuum image of SMMJ02399 from ALMA Cycle 0 – Band 9 Observations. The ALMA primary beam size (dashed-line) and synthesized beam (filled yellow) are shown. Only L1 (point-source component to the right) and L2SW (extended component to the left) are detected by ALMA.

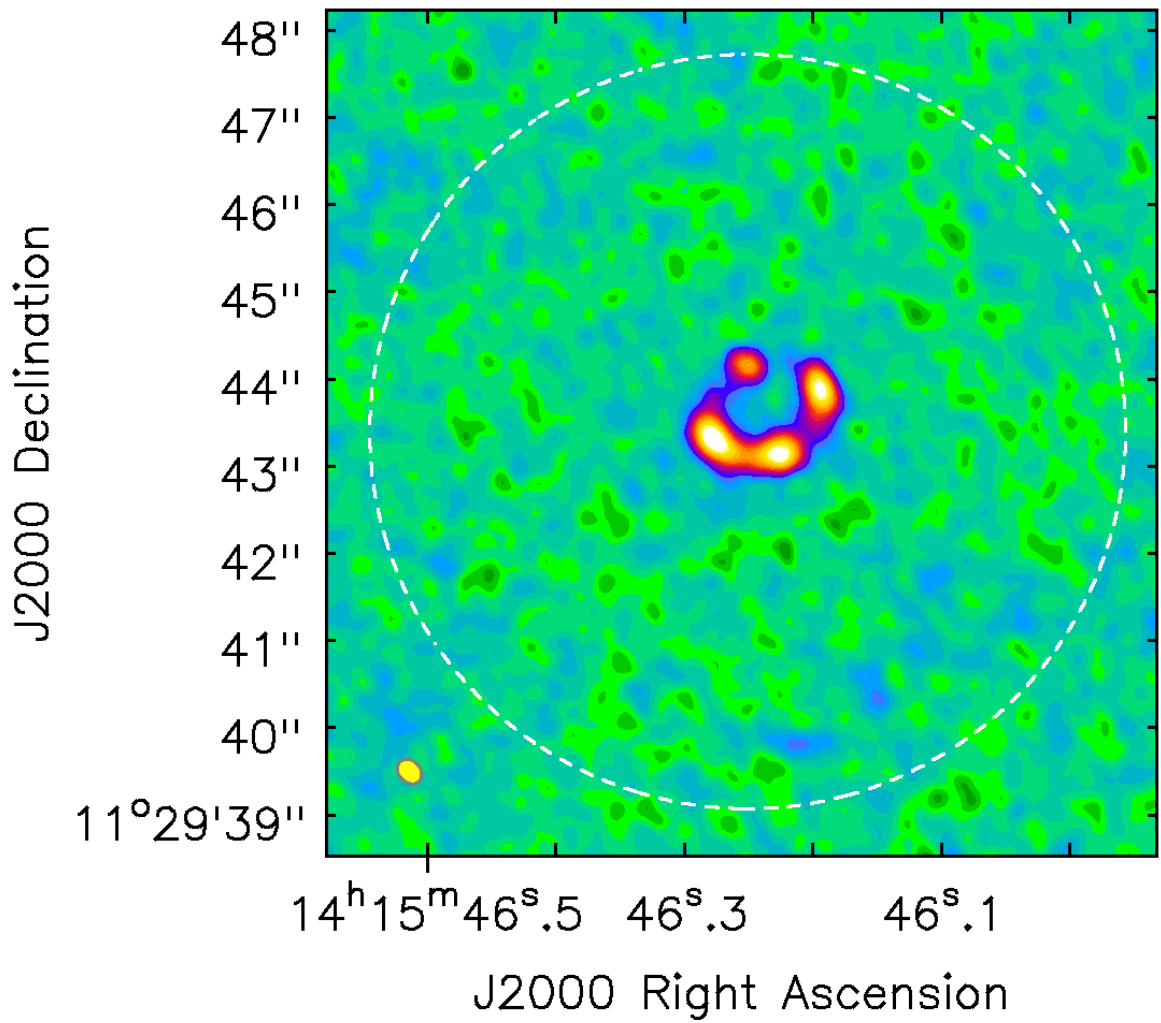

Figure 2: 122 μm rest frame continuum image of the Cloverleaf QSO from ALMA Cycle 0 – Band 9 Observations. The ALMA primary beam size (dashed-line) and synthesized beam (filled yellow circle) are shown.

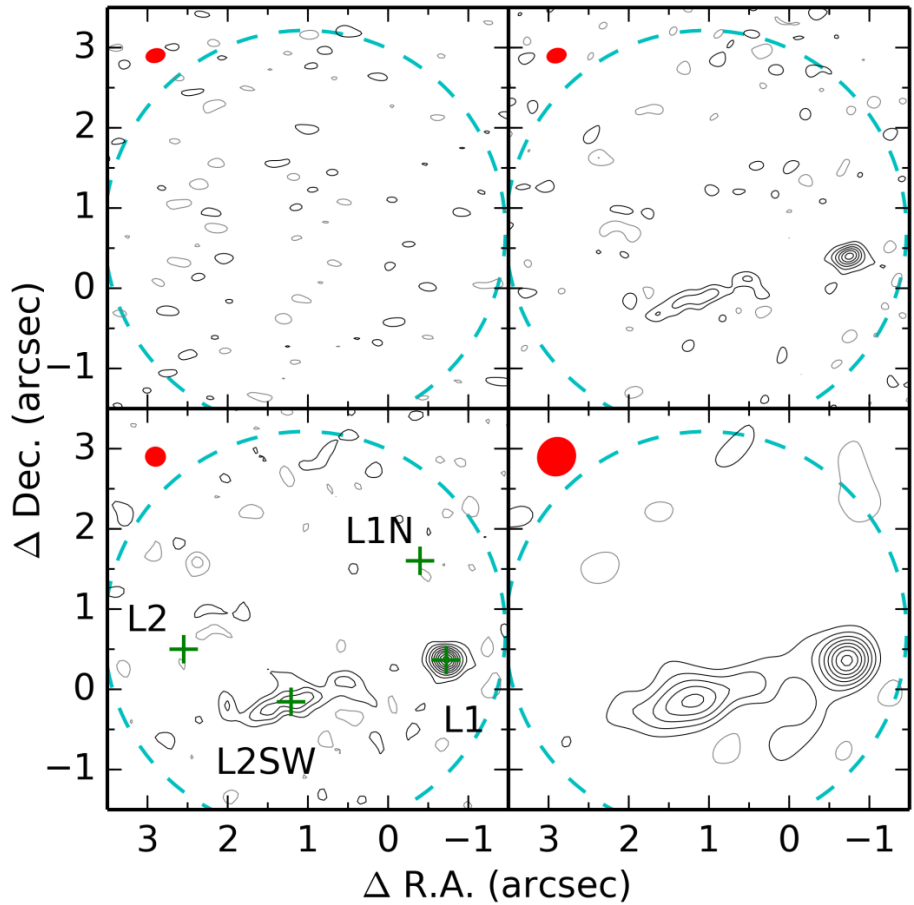

Figure 3: Continuum images of SMMJ02399 with uniform (upper left), natural (lower left), robust (upper right) and UV tapered to 0.5" (lower right) weighting schemes. Contour levels are at -2σ in grey and 2, 4, 6, 8, 10, 12, 14, 16, 18, and 20σ in black. The 1σ levels are 9.5, 1.1, 1.5, and 1.5 mJy beam$^{-1}$ respectively. The location of the different components (green crosses; based on the location of the peak emission in near-IR HST imaging) as well as the size of the synthesized beam (red) are shown. The cyan dashed-line indicates the aperture used for the full source-integrated flux-densities.

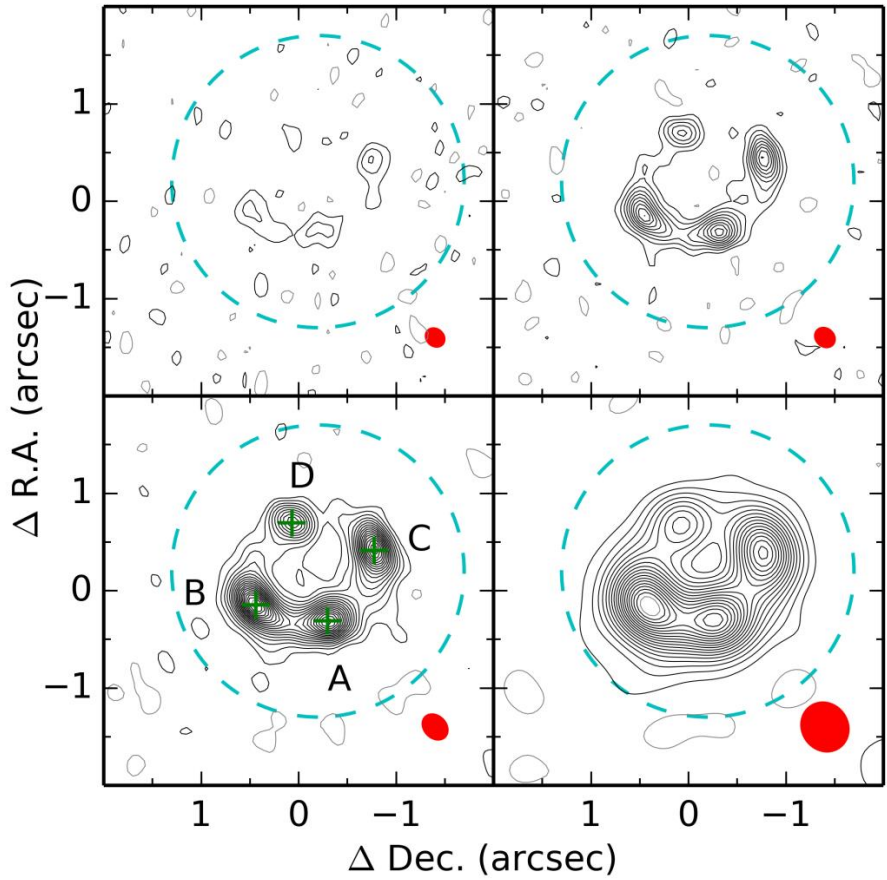

Figure 4: Continuum images of the Cloverleaf with Uniform (upper left), natural (lower left), Robust (upper right) and UV tapered to 0.5" (lower right) weighting schemes. Contour levels are at -2σ in grey and 2, 4, 6, 8, 10, 12σ, etc. up to 38σ in black. The 1σ levels are 4.4, 0.9, 1.2, and 1.5 mJy beam$^{-1}$ respectively. The location of the different components (green crosses) as well as the size of the synthesized beam (red) are shown. The cyan dashed-line indicates the aperture used for the full source-integrated flux-densities.

*Table 2: ALMA – rest frame 122 μm continuum properties*

| Component | Peak Intensity for Different UV Weighting Schemes | | | | Integrated Flux Density for Different UV Weighting Schemes | | | |
|---|---|---|---|---|---|---|---|---|
| (1) | (2) | (3) | (4) | (5) | (6) | (7) | (8) | (9) |
| | Uniform | Robust (R=0.0) | Natural | UV Tapered (0.5") | Uniform | Robust (R=0.0) | Natural | UV Tapered (0.5") |
| | mJy beam$^{-1}$ | mJy beam$^{-1}$ | mJy beam$^{-1}$ | mJy beam$^{-1}$ | mJy | mJy | mJy | mJy |
| SMMJ02399 | | | | | | | | |
| Beam Size (") | 0.25 x 0.18 | 0.26 x 0.19 | 0.26 x 0.25 | 0.50 x 0.48 | 0.25 x 0.18 | 0.26 x 0.19 | 0.26 x 0.25 | 0.50 x 0.48 |
| L1 | … | 20.2 | 20.7 | 25.7 | < 33.5 | 28.8 ± 5.3 | 29.2 ± 4.3 | 25.8 ± 2.4 |
| L2SW | … | 8.5 | 5.7 | 16.3 | < 43.3 | 36.8 ± 6.8 | 40.2 ± 4.3 | 33.2 ± 3.1 |
| L2 | … | 4.2 | 2.8 | 1.8 | < 51.7 | < 24.5 | < 15.3 | < 11.0 |
| Entire Source | … | … | … | … | < 54.8 | 65.5 ± 8.6 | 69.4 ± 6.1 | 59.0 ± 3.9 |
| RMS | 9.5 | 1.5 | 1.1 | 1.5 | … | … | … | … |
| The Cloverleaf | | | | | | | | |
| Beam Size (") | 0.23 x 0.19 | 0.24 x 0.20 | 0.31 x 0.25 | 0.54 x 0.49 | 0.23 x 0.19 | 0.24 x 0.20 | 0.31 x 0.25 | 0.54 x 0.49 |
| A | 18.6 | 23.5 | 31.7 | 52.5 | 30.9 ± 9.0 | 38.5 ± 2.4 | 42.0 ± 1.4 | 27.6 ± 1.2 |
| B | 18.9 | 24.8 | 36.0 | 59.6 | 35.2 ± 9.0 | 44.4 ± 2.4 | 48.9 ± 1.4 | 32.6 ± 1.2 |
| C | 26.9 | 22.4 | 29.8 | 45.5 | 33.1 ± 9.0 | 35.7 ± 2.4 | 38.2 ± 1.4 | 23.9 ± 1.2 |
| D | 16.1 | 13.3 | 18.0 | 25.0 | 12.0 ± 9.0 | 18.5 ± 2.4 | 21.3 ± 1.4 | 13.0 ± 1.2 |
| Entire Source | … | … | … | … | 80.0 ± 50.0 | 176.7 ± 13.9 | 260.3 ± 8.2 | 265.3 ± 7.2 |
| RMS | 4.5 | 1.2 | 0.9 | 1.5 | … | … | … | … |

Notes: All upper-limits given are 3σ limits

*3.2    Spectral Imaging*

   To produce the spectral data cubes for imaging of the [NII] 122 μm line we subtracted the continuum emission in both sources by fitting a first order polynomial to the continuum in the UV plane. The fit was performed over all channels excluding the range the [NII] line was expected to fall based on our ZEUS-1 observations. Additionally, we ignored the ~10 channels centered on any weak telluric feature. Once the continuum was subtracted we produced a dirty-image to estimate the 1-sigma RMS error. We then produced a cleaned spectral cube by cleaning down to a 3-sigma RMS level. For the Cloverleaf, a natural weighting scheme was sufficient, while for SMMJ02399 we imaged with a UV-taper to produce ~0.5" beam since the larger synthesized beam improved the signal-to-noise in the final extracted spectrum.

   In Figures 5 and 6 we show the extracted spectra from SMMJ02399 and the Cloverleaf. The regions over which the spectra are extracted are indicated on maps of the 122 μm continuum (grey-scale) along with the velocity integrated line flux (contours). For comparison we have also included the ZEUS-1 spectra for each source, scaled appropriately for clarity. Figures 7 and 8 show the ALMA spectrum for the individual source components of SMMJ02399 and the Cloverleaf respectively while the integrated line properties are listed in Table 3. In both sources the [NII] line is detected, at the 4σ level for SMMJ02399 and the 5σ level for the Cloverleaf. The line flux in the L2 component of SMMJ02399, the brightest component in the line, is $22 \pm 6$ Jy km s$^{-1}$ while the total flux, integrated over all components, in the Cloverleaf is $28 \pm 6$ Jy km s$^{-1}$. These are both are significantly weaker than observed with ZEUS-1. The line fluxes obtained with ALMA are ~17% and ~26% of the ZEUS-1 line fluxes for SMMJ02399 and the Cloverleaf respectively.

   The significantly weaker [NII] line flux as observed with ALMA can be explained if the line emission has been resolved out. We originally sought synthesized-beam resolutions of 0.7" and 0.8" FWHM for observations of the SMMJ02399 and the Cloverleaf respectively that would have given SNR at the line peak of 10 for SMMJ02399 and 5 for the Cloverleaf. These estimates assumed that the observed ZEUS-1 [NII] line fluxes were uniformly distributed in a circular region 1" in diameter in SMMJ02399 and 2" in diameter for the Cloverleaf, both reasonable given the knowledge of the sources at the time of our ALMA proposal in 2010. However, the final synthesized beams of ~0.25" FWHM give significantly more beams over the sources, which spreads the line emission into ~8 times more beams for SMMJ02399 and ~10 times more beams for the Cloverleaf. This suggests that we do not expect the ALMA detection of the [NII] emission in either source to be significant. Even if we assume the missing [NII] emission in the Cloverleaf is co-spatial with the ALMA continuum emission, which is extended over an annulus with inner and outer diameters of 1" and 2" respectively, the line emission would be split into ~7 more beams than expected so that the [NII] emission is indeed resolved out by the ALMA observations. Understanding SMMJ02399 is slightly more complicated. If the [NII] line is co-spatial with the ALMA continuum emission, a ~0.25" FWHM point source and ~0.25" x 1" extended source, which are fully recovered by ALMA, then the line would only be spread between approximately four or five of the 0.25" ALMA beams and should be strongly detected. For the [NII] line flux to be resolved out in SMMJ02399 we require that most of the [NII] emission occur on scales larger than the continuum emission. Our evidence supporting this scenario will be discussed in Section 4.1 below.

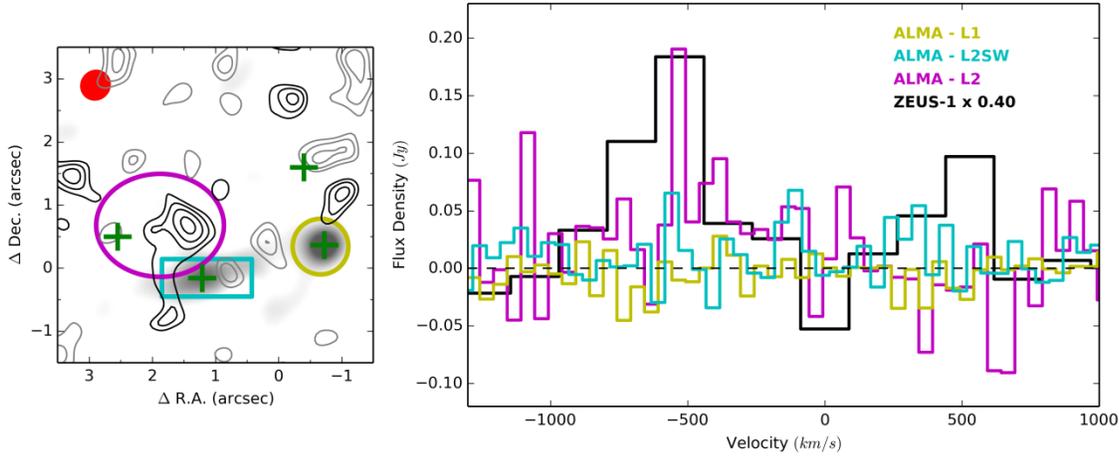

Figure 5: (left) SMMJ02399 velocity integrated [NII] line contours overlaid on the ALMA continuum image (gray-scale) with levels at -2.5, -2, -1.5 (gray), 1.5, 2, 2.5 and 3σ (black). The 1σ level equals 1.6 Jy km s$^{-1}$ beam$^{-1}$. Both the line and continuum maps have a ~0.5" synthesized beam as indicated in red. Regions used to extract spectra are indicated. L2, L2SW and L1 are respectively magenta, cyan, and yellow. As in Figure 3, the green crosses denote the location of the various components. (right) Spectrum extracted from the three regions with 50 km/s resolution in corresponding colors. The ZEUS-1 spectrum scaled by 0.4 is in black with Δv ~ 300 km/s.

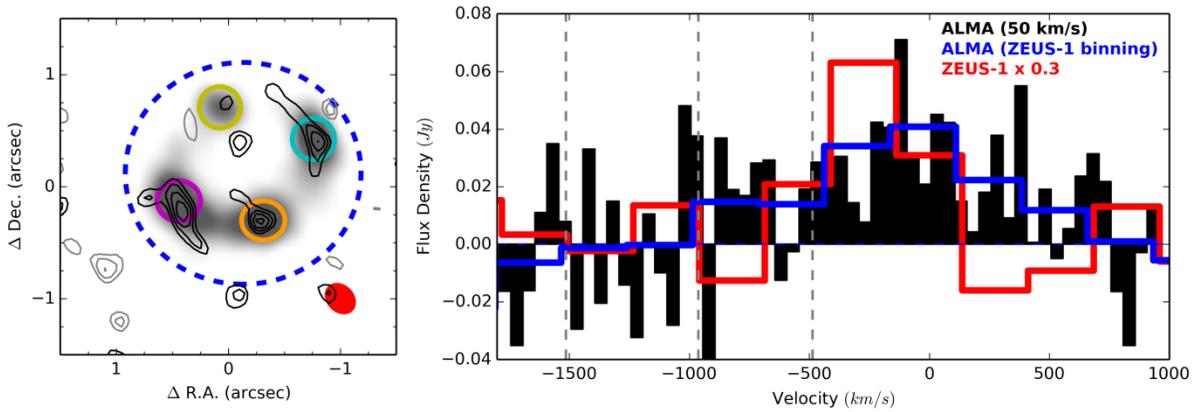

Figure 6: (left) Cloverleaf velocity integrated [NII] line contours plotted over the 122 μm continuum image (gray-scale) with levels at -3, -2.5, -2 (grey), 2, 2.5, 3 and 3.5σ (black). The 1σ level equals 1.5 Jy km s$^{-1}$ beam$^{-1}$. The regions used to extract spectra are indicated. The full source region is shown by the dashed blue circle while components A, B, C and D are indicated by orange, magenta, cyan, and yellow circles respectively. The synthesized beam of ~0.25" is shown in red. (right) Spectrum extracted over the entire source with 50 km/s resolution in black. The ZEUS-1 spectrum, scaled by 0.3, is shown in red while the ALMA spectrum with similar spectral binning (~ 300 km/s) is in blue. Vertical dashed lines indicate the location of telluric features that double the system temperature at those velocities.

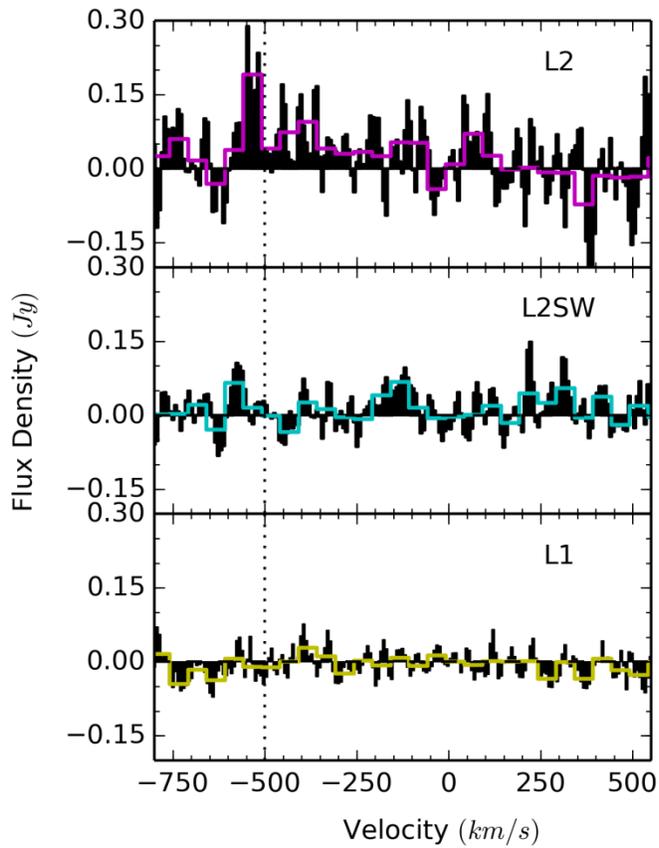

Figure 7: SMMJ02399 - ALMA spectra for the regions indicated in Figure 5. The filled black spectra are at the native ALMA spectral resolution of ~7.25 km s$^{-1}$, while the colored spectra have been binned to 50 km/s. A vertical dotted line at -500 km s$^{-1}$ indicates the expected velocity of the [NII] line based on the ZEUS-1 observations.

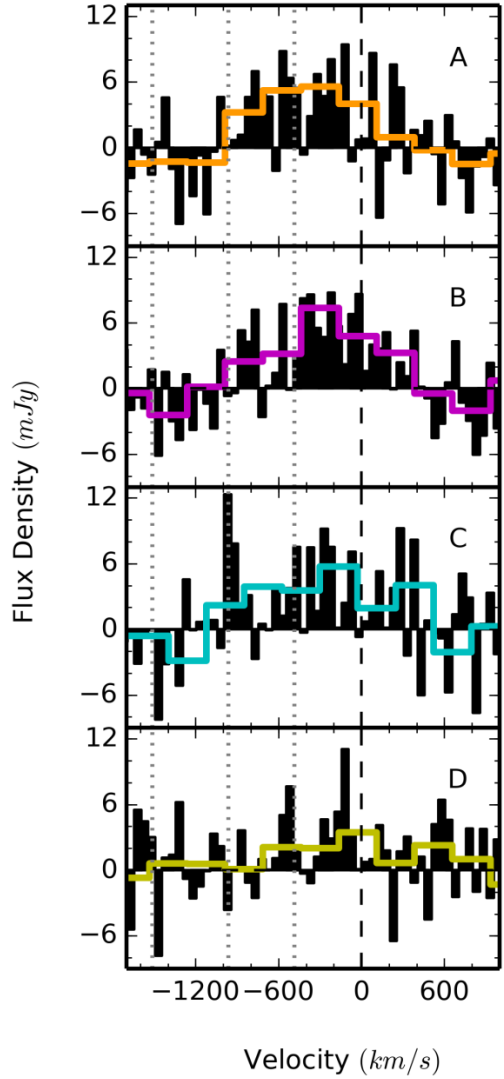

Figure 8: Cloverleaf-ALMA spectra for the regions indicated in Figure 6. The filled black spectra are binned to 50 km/s spectral resolution, while the colored spectra have the same spectral resolution as the ZEUS spectrum (i.e. $\Delta v \sim 300$ km/s). Vertical grey dotted lines indicate the location of weak telluric features while the black dashed line is at 0 km s$^{-1}$.

*Table 3: ALMA – [NII] Line fluxes*

| Source/Components | Line Flux | FWZF | Line Center |
|---|---|---|---|
| (1) | (2) | (3) | (4) |
|  | Jy km s$^{-1}$ | km s$^{-1}$ | km s$^{-1}$ |
| SMM J02399 | | | |
| L1 | < 6.0 [a] | 250 [b] | … |
| L2SW | 6.2 ± 2.1 | 150 | -130 |
| L2 | 21.9 ± 5.4 | 250 | -480 |
| The Cloverleaf | | | |
| A | 3.0 ± 0.9 | 1000 | 0 |
| B | 4.2 ± 0.8 | 1000 | 0 |
| C | 3.2 ± 0.9 | 1000 | 0 |
| D | 1.5 ± 0.8 | 1000 | 0 |
| Total | 27.8 ± 5.6 | 1000 | 0 |

Notes: Column 3 – FWZF is the full line width to zero flux used for integrating the spectra and determining the line flux. The values for SMMJ02399 are measured from the ALMA spectra. The Cloverleaf's widths are based on the high signal-to-noise observations of the CO(3-2) line (Weiß et al. 2003), which has width to zero flux of ~1000 km s$^{-1}$ and line center of 0 km s$^{-1}$. This is done since the low signal-to-noise and the week telluric features in the ALMA spectrum make it difficult to accurately determine directly.

[a] 3σ upper limit
[b] The line is undetected in this component, so the FWZF of the L2 component is assumed to obtain an upper limit to the line flux.

## 4. Nature of the [NII] emission

### 4.1 SMMJ02399

The ZEUS-1 spectrum from SMMJ02399 shows a line at -500 km s$^{-1}$ as well as a hint (~3σ) of another velocity component at 250 km s$^{-1}$ compared to the sources systemic redshift of z~2.8076 that is based on CO(3-2) observations (Genzel et al. 2003). In our previous work (Ferkinhoff et al. 2011) we associate the blue velocity component of the [NII] emission with L2SW and the red velocity component with L1. We base this on long-slit optical spectroscopy and low spatial resolution millimeter spectroscopy in the literature (Ivison et al. 1998, 2010; Genzel et al 2003) that showed blue and red velocity components in both Lyα and CO(3-2) were spatially coincident with L2SW and L1 respectively. Making L2SW the source of the [NII] emission is consistent with the [NII] line being produced in a star-formation paradigm, since L2SW is the site of an extreme and dusty starburst. Given the quality of the previous spectroscopic data, however, these associations are not conclusive. The previous observations lack the resolution, spatial and spectral, necessary to truly separate the L2 and L2SW components. It is not possible to completely rule-out that some part of the [NII] emission detected with ZEUS-1 comes from L2 or even the L1N component. Our ALMA data here are the first observations with high enough spatial and spectral resolution to clearly identify from which component the emission arises.

The ALMA observations clearly detect the blue component (Figure 7) and surprisingly it is not spatially coincident with any strong dust continuum emission. In other words the strongest line at -500 km s$^{-1}$ comes from the L2 component, not the dusty starburst in L2SW or the quasar in L1—the sources of the strong dust continuum emission. Albeit, the blue component that is recovered by the ALMA [NII] observations is not at the peak of the L2 emission from the optical-HST imaging, (see Fig. 9) but is none the less coincident with significant emission. Furthermore, since the ALMA detection of the [NII] line is at low signal to noise and is missing significant flux, this offset between the peaks in the optical and [NII] emission should not be considered significant. In Figure 7 we also show the source integrated [NII] lines for the L2SW and L1 components. No line is detected from L1, while L2SW shows a hint (~3σ) of a line at -130 km s$^{-1}$. Since this only a marginal detection, and too weak to been seen in the ZEUS-1 spectrum, consider only the ALMA line flux from L2 when comparing to the ZEUS-1 [NII] line. In this case, as with the Cloverleaf, the majority (~86%) of the ZEUS-1 velocity integrated line flux in SMMJ02399 remains undetected. If the unrecovered line flux is coincident with L2 (and perhaps L1N), then it must be extend over ≥ 1 square-arcsecond to be undetected in our ALMA observation given their noise level.

The significant discrepancy between the observed ALMA and ZEUS-1 line fluxes from our sources is reconciled if the [NII] line emission-regions are extended so that the flux is resolved out. It could also be that either ALMA or the ZEUS-1 observations are poorly calibrated. We have no reason to doubt the ALMA calibration however, since as described in Section 2.1, we correctly recover the fluxes of the bandpass and phase calibrators. An independent check on the validity of the ZEUS-1 calibration is achieved by comparing the observed [NII] line fluxes to those expected given an observed thermal free-free radio continuum flux that would be produced in the same HII regions as the [NII] line. Fortunately, radio continuum observations from SMMJ02399 are available in the literature to enable this comparison. The flux-density of SMMJ02399 has been observed with the JVLA (Jansky Very Large Array) at 1.4, 5 and 32 GHz to be 526 ± 50, 130 ± 22 and 57 ± 25 µJy respectively, at sufficiently large enough beams (≳ 2") to ensure no flux has been resolved-out (Thomson et al. 2012, Ivison et al. 1998). These correspond to the rest frame continuum frequencies 5.3, 19, and 122 GHz so that they probe the free-free continuum produced in stellar HII regions. While the frequency bands will be contaminated by non-thermal synchrotron emission at low frequencies and dust emission at high frequencies, the flux densities can still provide an estimate of the maximum [NII] line flux that we can reasonably expect from SMMJ02399. The specific emissivity for free-free emission is (see Rybicki & Lightman 1986)

$$\varepsilon_\nu^{ff} = 6.8 \times 10^{-38} Z^2 n_e n_i T^{-1/2} e^{-h\nu/kT} \bar{g}_{ff}. \qquad (1)$$

Since free-free emission is dominated by electron/proton interactions so $Z \cong 1$ (the charge number) and $n_e \cong n_i$ (the electron and ion number densities). T, $\nu$, and $\bar{g}_{ff}$ are respectively the gas temperature in which the free-free emission arises, the frequency at which the free-free emission is observed, and the free-free gaunt-factor at those same frequencies. The specific luminosity for free-free emission is the volume integral over this specific emissivity,

$$L_\nu^{ff} = \int_V \varepsilon_\nu^{ff} dV \cong \varepsilon_\nu^{ff} \cdot V, \qquad (2)$$

and the [NII] line to free-free specific luminosity ratio in the low density limit of the [NII] line is then

$$\frac{L_{[NII]}}{L_\nu^{ff}} = \frac{h\nu_{21} n_e n_{N^+} q_{02} \cdot V}{\varepsilon_\nu^{ff} \cdot V} = \frac{3.88 \times 10^{-22} \frac{n_{N^+}}{n_e} n_e^2}{6.8 \times 10^{-38} n_e^2 T^{-1/2} e^{-h\nu/kT} \bar{g}_{ff}} \quad [Hz], \qquad (3)$$

where $q_{02}$ is the collisional rate coefficient between the $^3P_0$ and $^3P_2$ levels of ionized-nitrogen, $\nu_{21}$ is the frequency of the [NII] line, $n_{N^+}/n_e$ is the singly-ionized nitrogen abundance, and V is the volume of emitting gas. Assuming T = 8000 K, we have $q_{02} = 2.365 \times 10^{-8}$ cm$^3$ s$^{-1}$ and $\bar{g}_{ff}$ = 4.86, 4.18, and 3.34 for frequencies, $\nu$, of 5.3, 19, and 122 GHz[3]. We pick T = 8000 K based on our HII region modeling in Ferkinhoff et al. 2011 that suggests the [NII] emission can arise from HII regions with temperatures between 6000 and 10000 K. The expected [NII] 122 μm line flux in terms of the free-free flux-density, $S_\nu^{ff}$, is then

$$F_{[NII]} \cong S_\nu^{ff} \cdot 5.10 \times 10^{17} \frac{n_{N^+}}{n_e} / \bar{g}_{ff} \quad [W/m^2], \qquad (4)$$

where $S_\nu^{ff}$ is in units of W m$^{-2}$ Hz$^{-1}$. Since $\exp[(-h\nu)/kT]$ in equation 3 is ~1 for the three radio frequencies we consider, equation 4 can be used for all three frequencies. To estimate our expected [NII] line flux using equation 4, we must assume a singly-ionized nitrogen to hydrogen abundance. For simplicity, we assume all of the nitrogen in the HII region is singly ionized and that the total nitrogen to hydrogen abundance is the same as the Milky Way gas phase abundance (Savage & Sembach 1996; $n_N/n_e$ = 9.3 × 10$^{-5}$). The latter assumption is reasonable considering we do not have a direct measure of the metallicity within SMMJ02399 and studies show dusty high-z systems can span a range of metallicities from low-metallicity to significantly more metal rich than the Milky Way (Casey, Narayanan & Cooray 2014). The former assumption makes the free-free estimate of the [NII] line flux an upper limit given the assumed N/H ratio.

Given the continuum observations at 1.4, 5 and 32 GHz and using Milky Way abundances, we would expect maximum [NII] line fluxes of 5.1 × 10$^{-17}$, 1.3 × 10$^{-17}$ and 8.1 × 10$^{-18}$ W m$^{-2}$. For comparison, the [NII] line flux observed ZEUS-1 is (2.77 ± 0.43) × 10$^{-18}$ W m$^{-2}$. The continuum at 1.4 and 5 GHz may have significant contributions from synchrotron emission, so the [NII] line flux is likely over estimated by these observations. The estimate based on the 32 GHz continuum on the other hand is only ~3 times higher than observed by ZEUS-1. Thomson et al. (2012) calculate that about half the total continuum emission at 32 GHz is produced by free-free emission (24 ± 6 μJy), with the rest coming from synchrotron and dust emission, in which case the free-free estimate agrees significantly, (3.4 ± 0.9) × 10$^{-18}$ W m$^{-2}$, with the flux observed by ZEUS-1.

This estimate is still subject to our assumptions about the singly-ionized nitrogen to hydrogen abundance, so that it is only upper limit and also consistent with the ALMA [NII] line flux as well. However, we also observed the [OIII] 88 μm line from SMMJ02399 so we can constrain the ionization state of the HII regions—assuming the [OIII] emission arises from the same regions as the [NII]—and reduce our assumptions to metallicity effects only. Based on the models of the [OIII] and [NII] emission in Ferkinhoff et al. (2011) we expect ~74% of the nitrogen to be singly ionized so that the best estimate of the [NII] emission via free-free emission is $(2.6 \pm 0.6) \times 10^{-18}$ W m$^{-2}$, which is excellent agreement with the ZEUS-1 value. Further, it is consistent with our ALMA observations only if the gas phase abundance in SMMJ02399 is ~5x lower than seen in the Milky Way. Lastly, the observed-frame 1.4 GHz continuum, as shown in Figure 9, extends over most of the source including the L2 component. With our assumption that the free-free and [NII] emission should be co-spatial, then we expect the [NII] line emission to be extended as well. This work strongly suggests that the ZEUS-1 [NII] line flux is valid, and we are missing a significant amount of flux in our ALMA observations.

---

[3] The collisional rate coefficient, $q_{02}$, is calculated for a gas temperature of 8000 K using the collisional strengths from Hudson & Bell 2004.

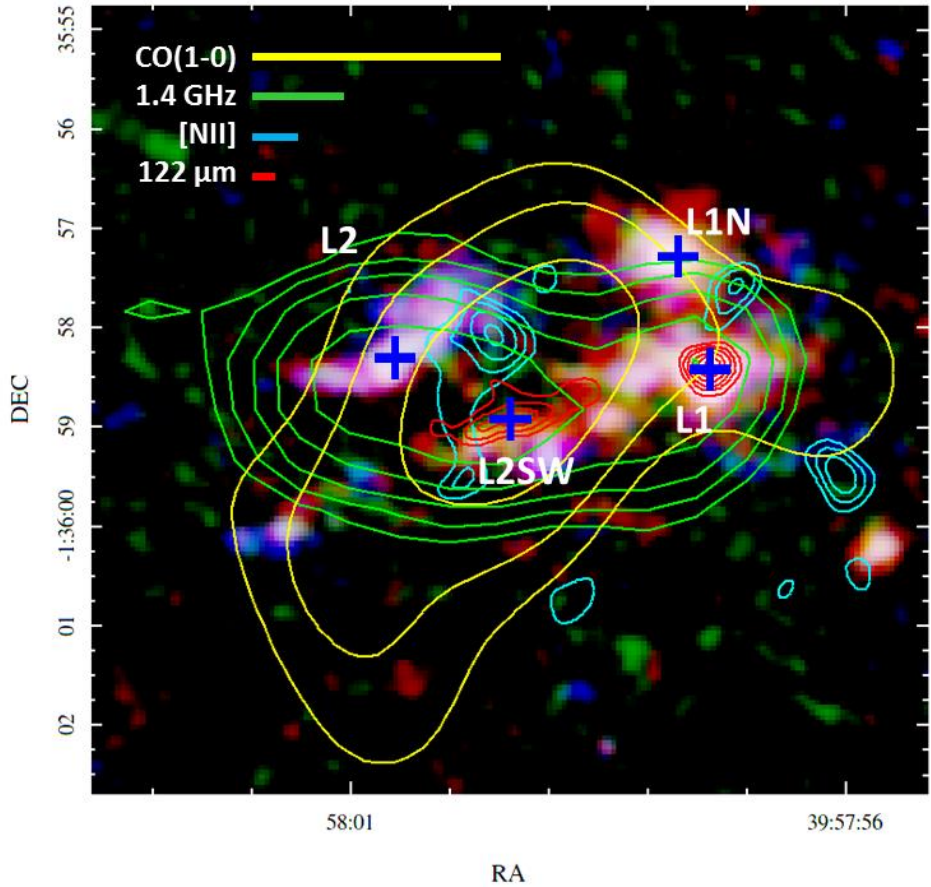

Figure 9: ALMA 122 μm rest-frame continuum contours (red) and the integrated [NII] 122 μm line (cyan) of SMMJ02399 with contours of the observed-frame 1.4 GHz continuum (green; Ivison et al. 2010, Owen & Morrison 2008) and integrated CO(1-0) (yellow; Ivison et al. 2010, Thomson et al. 2014) emission overlaid on a RGB composite of the F475W (*g*), F675W (*r*), and F814W (*i*) HST bands (Ivison et al 2010, Aguirre et al. 2013). The peaks for each component in the HST imaging are noted with a blue cross. L1sb, as identified in the HST imaging by Aguirre et al. (2013), is the red emission in the HST image in the bottom half of the source. It encompass both of the ALMA continuum sources show in red at L2SW and L1. In the upper-left corner we note the approximate size of the synthesized beam (FWHM) for each set of contours.

*4.2    The Cloverleaf*

The [NII] line emission from the Cloverleaf that is detected by ALMA is primarily coincident with the location of the multiply-lensed quasar point source, which is seen in the velocity integrated map (Figure 6). Simply summing the spectra obtained from the positions of the quasar images (Figure 8) recovers the bulk of the source integrated flux shown in Figure 6. Furthermore, the line emission appears unresolved with the ALMA beam of ~0.25" corresponding to ~180 pc in the source plane. As such we attribute the [NII] line detected by ALMA to be primarily produced in the NLR of the Cloverleaf's AGN, though may be contaminated with some emission due to star formation. The ZEUS-1 line flux then is the total line strength from the NLR and star formation combined. A lower limit on the contribution to the line from star formation is the difference between the ZEUS-1 and ALMA values. Specifically, given that only ~20% of the ZEUS-1 velocity integrated line flux is recovered by ALMA, the remaining ≥80% is likely attributed to extended emission from star formation in the quasar host.

## 5.    Analysis

Using the line fluxes reported in Table 3 and the continuum flux-densities in Table 2 we derive several intrinsic source properties. The derived properties related to star formation in each source are summarized in Table 4. All values have been de-magnified assuming the magnification factors in Table 1. Given that the SMMJ02399 is weakly lensed by a foreground cluster, we apply the same magnification factor to each component. The Cloverleaf on the other hand is strongly lensed so accounting for the magnification is more complicated. First, the various components of the [NII] line and 122 µm continuum seen in our ALMA observations (e.g. A,B, C and D) are images of the same source. As such we do not consider the individual components of the Cloverleaf in our calculations below. Instead we consider only the "total" source integrated [NII] and continuum emission. Further, because the ALMA detected [NII] line is likely associated with the Cloverleaf's AGN of the Cloverleaf quasar we don't use it directly. Instead we consider the [NII] emission associated with star formation to be the ZEUS-1 line flux *minus* the ALMA line flux. We use this line flux (~80 Jy km s$^{-1}$, i.e. the missing ALMA flux) combined with a magnification factor of 11 to determine the intrinsic properties of the Cloverleaf's starburst (e.g., star formation rate, ionizing flux, etc.) in the analysis below.

The magnification we employ for the Cloverleaf comes from the lens and source modeling of Venturini & Solomon (2003) who simultaneously model the source and gravitational lens based on the observed CO(7-6) map. They determine that the CO(7-6) emission arises from a disk with FWHM of ~1.5 kpc and likely associated with star formation, and not the central engine. It is not immediately obvious that applying this magnification in our analysis is appropriate. By looking however at the four lensed components (A,B,C and D) in our 122 µm map we see the positions and relative strengths of their peak flux densities is identical to the components of the CO(7-6) map in Venturini & Solomon (2003). This suggest that both the CO(7-6) and 122 µm emission likely experience similar gravitational lensing, and applying the Venturini & Solomon magnification to our ALMA continuum observations is reasonable. Conversely, applying this magnification directly to our ALMA [NII] line flux is likely inappropriate. If this [NII] emission arises from the NLR of the AGN it would have a different distribution in the source plane than that due to star formation and subsequently a different magnification due to gravitational lensing. Indeed, Chae & Turnshek (1999) model the lensing of the Cloverleaf's quasar emission and determine a magnification of ~18. However, in our analysis below we consider the [NII] emission arising from the Cloverleaf's star formation, i.e. the missing [NII] flux in our ALMA observations, which we have argued above is likely coincident with our 122 µm continuum emission. As such, applying the Venturini & Solomon (2003) magnification to the star-formation associated [NII] line-flux is also reasonable. Of course to estimate the Cloverleaf's intrinsic [NII] emission due to star formation we subtract the ALMA [NII] line flux from our ZEUS-1 flux, so that by taking this difference and applying the Venturini & Solomon magnification this estimate should be considered a strong a lower limit in the event that the AGN associated [NII] emission is indeed more strongly lensed as suggested by the Chae & Turnshek models.

## 5.1 [NII] Line Diagnostics: N/H Abundance, Ionized Gas Mass, Ionizing Flux, Mass of Ionizing Stars and Star formation Rate.

Based on our work in section 4.1 we can take the ZEUS-1 line flux from SMMJ02399 to be the correct measure of the total [NII] line flux in the system and determine the nitrogen abundance of SMMJ02399 directly, instead of assuming a Milky Way value. To do so, we write equation 4 so that the nitrogen abundance is a function of the [NII] and free-free emission. Using this equation, the [NII] line flux, the fraction of nitrogen that is singly ionized based on the [OIII]/[NII] ratio and the free-free emission at 32 GHz we determine SMMJ02399's nitrogen to hydrogen abundance ratio to be $(1.1 \pm 0.3) \times 10^{-4}$. This is consistent with a Milky Way abundance we assumed prior.

Following Ferkinhoff et al. 2011 we estimate the minimum amount of ionized gas needed to produce the observed ALMA line flux assuming the line is emitted in the high-density and high temperature limit for the transition (Table 4). This is a minimum mass since HII-region densities on galactic scales are typically much lower than the critical density (~ 310 cm$^{-3}$), and not all of the nitrogen in that region will be singly ionized. Regardless, the [NII] line observed with ALMA suggests the sources contain $>10^9$ M$_\odot$ of ionized gas.

On the other hand, if the [NII] line is emitted in the low-density and high-temperature limit, the [NII] line flux is then proportional to the hydrogen ionizing photon rate, $Q_0$, required to maintain the ionization equilibrium. The ionizing photon rate required to maintain the ionization in a of volume V is

$$Q_o = V \cdot n_e^2 \alpha_B \qquad (5)$$

Where $\alpha_B$ is hydrogen recombination rate to all excited (n>1) levels and $n_e$ is the electron number density in the region. We equate the electron density here to the electron density in the relation for the [NII] line luminosity in the low-density and high temperature limit (the numerator of Eq. 3) to get the ionizing photon rate as a function of the [NII] line luminosity, L$_{[NII]}$,

$$Q_o = \frac{\alpha_B}{q_{02} h \nu_{21} \frac{n_{N^+}}{n_e}} L_{[NII]} = 2.6 \times 10^{46} \left\{\frac{10^{-4}}{\chi_{N^+}}\right\}\left\{\frac{L_{[NII]}}{L_\odot}\right\} \; [sec^{-1}]. \qquad (6)$$

This assumes that all nitrogen in the ionized region is singly ionized such that $\chi_{N^+} = \chi_N = n_N/n_H$, the gas phase nitrogen to hydrogen abundance. Here we assume Milky Way gas phase metallicities ($9.3 \times 10^{-5}$; Savage & Sembach 1996), which is appropriate based on the abundance estimate above. Since nitrogen requires between 14.5 and 29.6 eV photons to be singly ionized, the ionizing photon rate as determined via [NII] arises mostly from stars between stellar types B1 and O7.5. Assuming a stellar model and using the ionizing rate of such stars (e.g Vacca et al. 1996), one can estimate the number and mass of stars of a specific spectral type required to produce the observed [NII] line luminosity.

Using Equation 6 we obtain an ionizing photon rate of $(11.7 \pm 2.9) \times 10^{55}$ and $(3.3 \pm 1.1) \times 10^{55}$ ph. sec$^{-1}$ for the L2 and L2SW components of SMMJ02399. This corresponds to $(4.0 - 28) \times 10^8$ and $(1.1 - 8.0) \times 10^8$ M$_\odot$ of stars respectively, where the range in mass is set by assuming only O7.5 stars or only B1 stars are the source of the ionizing photons. These numbers are reasonable given the stellar masses reported in Aguirre et al. (2013; M$_\star$ = $6.3 \times 10^{11}$ M$_\odot$ for L1 & L2SW combined and $1.3 \times 10^{10}$ M$_\odot$ for L2). For the Cloverleaf we obtain an ionizing photon rate of $(7.7 \pm 0.9) \times 10^{55}$ ph. sec$^{-1}$ that is produced by $(2.6 - 19) \times 10^8$ M$_\odot$ of stars by using the [NII] line-flux from star formation only as described above.

Constraining the ionizing photon rate, $Q_0$, allows us to determine the star formation rate of the sources. We assume the star-formation rate indicator for $Q_0$ from Calzetti (2007),

$$SFR(Q_0)[M_\odot \, yr^{-1}] = 7.4 \times 10^{-54} \, Q_0 \, [sec^{-1}], \qquad (7)$$

which is calibrated assuming a Kroupa-IMF for stars down to 0.1 $M_\odot$ and continuous star-formation over the past 100 Myr. We obtain star-formation rates of 870 ± 210 and 240 ± 80 $M_\odot$ yr$^{-1}$ for L2 and L2SW in SMMJ02399 and 570 ± 70 $M_\odot$ yr$^{-1}$ in the Cloverleaf.

*5.2    FIR Luminosity, Dust Mass, and Obscured Star-Formation Rate.*

For both SMMJ02399 and the Cloverleaf there are no observations of their rest frame FIR continuum at different wavelengths that have spatial resolutions comparable to our ALMA 122 μm continuum image. As such we are unable to perform detailed SED fitting of each component to determine their FIR luminosities. We work around this problem by using the component-to-total flux-density ratio for each component, as observed in our ALMA continuum observations, to scale the total FIR luminosity in Table 1 to the expected luminosity of each component. This is reasonable considering that our ALMA observations are probing the emission near the peak of the dust SED, providing a reasonable measure of a component's FIR emission. For L2 in SMMJ02399, where we do not detect the dust continuum with ALMA, we assume a 1σ upper limit for its flux density. This is likely an over estimate considering that the combined flux densities from L2SW and L1 in the ALMA observations fully recover the continuum flux-density of the entire source as measured by single-dish observations. Scaling the total FIR luminosity[4] by the fraction of the 122 μm rest-frame continuum flux-density for each component we obtain FIR luminosities of (2.3 ± 0.8) × 10$^{12}$ $L_\odot$ for L2SW and < 0.9 × 10$^{12}$ $L_\odot$ for L2 respectively. Similarly, the FIR luminosity in L1 is (1.7 ± 0.6) × 10$^{12}$ $L_\odot$. A caveat to this approach is if the components are dominated by emission with different dust temperatures, in which case the 122 μm rest-frame continuum flux-density of each component will actually scale to a FIR luminosity differently. That being said, our scaled FIR luminosity for L1 agrees with previous estimates of the AGN contribution to the total FIR luminosity (~50%) from Lutz et al. (2005), suggesting our approach is reasonable.

Using the ALMA 122 μm continuum flux density and assuming a dust temperature we can make a rough estimate of the dust mass necessary to produce the observed continuum emission. The mass of dust, $M_d$, is related to the observed specific luminosity, $L_\nu$, by,

$$L_\nu = 4\pi \kappa_d(\nu) B_\nu(T_d) M_d \qquad (8)$$

Where $\kappa_d(\nu)$ is the dust opacity per unit mass of dust, $B_\nu(T_d)$ is the Planck function at dust temperature $T_d$. Assuming a Li & Draine (2001) dust model[5], the dust masses in the L1 and L2SW components of the SMMJ02399 system are (3.0 ± 0.9) × 10$^8$ $M_\odot$ and (4.2 ± 1.3) × 10$^8$ $M_\odot$ respectively, assuming a dust temperature of 45 ± 5 K. The L2 component upper-limit is < 2.1× 10$^8$ $M_\odot$ equivalent to a dust surface-density of 65 $M_\odot$ pc$^{-2}$. This reproduces well previous SED modeling of SMMJ02399 that gave dust temperatures between 30 and 50 K with a dust mass of 7-5 × 10$^8$ $M_\odot$ (Ivison et al. 1998).

For the Cloverleaf we perform a two temperature SED fit with the Li & Draine (2001) dust model to all of the photometry in Weiß et al. (2003). Weiß et al. perform a similar fit, though they assume a simple power-low dust opacity relation. They attribute the warm dust component to the AGN and the cold dust component to star formation. In our fit we determine dust temperatures of 130 K and 33 K respectively for the warm and cold dust components, where the warm component is 1.52 ×10$^6$ $M_\odot$ and cold is 1.09 ×10$^9$ $M_\odot$. These dust masses are significantly different than those obtained by Weiß et al. (see Table 1). This is due to our use of the Li & Draine (2001) dust model, which more accurately describes the dust-opacity than the simple power-law used in Weiß et al., but subsequently requires a higher and lower dust temperature for the warm and cold components respectively to match the observed photometry. This in turn means our SED model gives less mass in the warm component and more mass in the cold component than found by Weiß et al.

---

[4] We define the FIR luminosities to be from 42.5 to 122.5 μm (covering the IRAS 60 and 100 μm bands) following the prescription of Helou et al. (1986). Some authors extend the FIR luminosity to include wavelengths up to 500 microns; these values are typically ~1.5 times larger than the luminosities we report here. The total infrared luminosity, $L_{IR}$, is the integrated luminosity between 8 and 1000 microns and is ~2 times larger than the FIR luminosity as we define above.

[5] We use dust opacities for the Li & Draine (2001) dust model from Weingartner & Draine (2001) that have been renormalized according to the discussion in Draine 2003. Use of these dust opacities here assumes that our sources have Milky Way like dust opacities. Given that we show SMMJ02399 has nitrogen abundance consistent with the Milky Way value, this assumption is reasonable.

Using the resolved continuum map and subsequent scaled FIR luminosities of the SMMJ02399 source components allows us to determine the star formation rate (SFR) in each component as traced by the star light that is absorbed by dust and re-emitted at infrared wavelengths: the dust-obscured star-formation. For many galaxies, the FIR star-formation indicator is a good measure of their total star formation rate. However, for a galaxy with very little dust it significantly underestimates the amount of star formation (Bell 2002). Using the Kennicutt (1998) FIR star formation indicator

$$SFR(L_{FIR})[M_\odot \, yr^{-1}] = 1.1 \times 10^{-10} \left[\frac{L_{FIR}}{L_\odot}\right] \, , \quad (9)$$

gives star formation rates of $300 \pm 100$, $390 \pm 130$ and $< 150$ $M_\odot$ yr$^{-1}$ for SMMJ02399 components L1, L2SW and L2 respectively. Similarly, using the total FIR luminosity of the Cloverleaf we estimate that it has a star formation rate of $950 \pm 30$ $M_\odot$ yr$^{-1}$.

*Table 4: Derived Star Formation Properties*

| (1) | $L_{[NII]}$ (2) $10^9 L_\odot$ | $M_{H+}$ (3) $10^9 M_\odot$ | $Q_0$ (4) $10^{55}$ sec$^{-1}$ | $N_{Stars,Q0}$ (5) $10^7$ | $M_{Stars,Q0}$ (6) $10^8 M_\odot$ | $L_{FIR}$ (7) $10^{12} L_\odot$ | $M_{dust}$ (8) $10^8 M_\odot$ | $SFR_{[NII]}$ (9) $10^2 M_\odot$ yr$^{-1}$ | $SFR_{FIR}$ (10) $10^2 M_\odot$ yr$^{-1}$ |
|---|---|---|---|---|---|---|---|---|---|
| | | | | SMM J02399 | | | | | |
| L2SW [a] | $0.98 \pm 0.33$ | $0.64 \pm 0.22$ | $3.3 \pm 1.1$ | $0.33 - 4.4$ | $1.1 - 8.0$ | $2.3 \pm 0.8$ | $4.20 \pm 1.3$ [c] | $2.3 \pm 0.8$ | $3.9 \pm 1.3$ |
| L2 [a] | $3.5 \pm 0.9$ | $2.3 \pm 0.4$ | $12.0 \pm 3.0$ | $1.2 - 15$ | $0.92 - 6.5$ | $< 0.90$ | $< 2.10$ | $8.7 \pm 2.1$ | $< 1.5$ |
| | | | | The Cloverleaf | | | | | |
| Total [b] | $2.3 \pm 0.6$ | $1.5 \pm 0.4$ | $7.7 \pm 0.9$ | $0.8 - 9.7$ | $2.6 - 19$ | $5.6 \pm 0.20$ [d] | $11.0 \pm 3.0$ [e] | $5.7 \pm 0.7$ | $9.5 \pm 0.3$ |

Columns: (1) component name, (2) [NII] line Luminosity, (3) ionized gas mass, (4) ionizing photon rate, (5) number of ionizing stars, (6) mass of ionizing stars, (7) FIR luminosity, (8) mass of dust, (9) SFR traced by the [NII] line, (10) SFR traced by the FIR luminosity.

Notes: [a] [NII] derived values for SMMJ02399 are based on the ALMA line fluxes for the respective component listed in Table 3. Since the ALMA observations resolved out significant flux, the derived values should be considered lower limits.
[b] [NII] derived values for the Cloverleaf use the ZEUS-1 observed line flux minus the total ALMA line flux from Table 3.
[c] Dust mass estimated assuming a single dust temperature of $45 \pm 5$
[d] Value from the literature as reported in Table 1
[e] Dust mass is for the cold, 33 K, SED component associated with star formation determined via the two-component SED fit in Section 5.2.

Looking closely at the results for SMMJ02399, one should consider its star formation rate in L1 cautiously since its dust emission is likely produced by the AGN and not star formation. The FIR derived star formation rate for L2SW agrees with the rate determined via its [NII] emission while the two rates for L2 significantly disagree. The [NII] derived SFR is $\geq 6$ times larger in L2 than suggested by the upper limit of its FIR luminosity. This disagreement can be explained if L2 is the site of significant unobscured star formation. The star formation rate as traced by the [NII] line should probe both obscured and unobscured star formation, so that in L2SW—which is heavily obscured by dust—the FIR and [NII] derived SFR agree with each other. In the L2 component, where there is very little dusts the FIR derived SFR significantly underestimates the amount of star formation present, which is correctly traced by the [NII] line. The high SFR in L2 as traced by the [NII] line, the missing [NII] flux in the ALMA observation that must be coming from scales larger than the 122 μm continuum emission in L2SW, and the fact that L2 and L1N show no 122 μm continuum emission suggest that SMMJ02399 must have significant amount of unobscured star formation.

## 6. Discussion

The goal of our ALMA observations of the Cloverleaf and SMMJ02399 was to map the [NII] emission and determine the contribution from the individual components to the line emission. While the data does not provide detailed maps of the [NII] line, and in fact only recovers a fraction of the total [NII] line flux observed by ZEUS-1, the data is still quite powerful and allows us to gain new insights into both sources. In Figure 10 we plot the [NII] line luminosity to FIR luminosity ratio as a function of the FIR luminosity for SMMJ02399, the Cloverleaf and other nearby and high-z systems. We have included the ZEUS-1 observed values, as well as the ALMA line luminosities. The ZEUS-1 value for the Cloverleaf is fairly typical compared to other systems while the ratio for SMMJ02399 is at the upper range of values seen in other systems, though certainly not the most extreme. Examining the values of the individual components in SMMJ02399 obtained through our ALMA program paints a very interesting picture. The line-to-continuum ratio for L2SW (the cyan square) and the L1 upper limit (yellow square) are in the range of typical values seen in nearby systems. In L2, however, the strong line-emission in this area, and undetected dust continuum gives a lower limit for the line to continuum ratio of > 0.004. This is ratio is higher than that seen in all but one source, the brightest cluster galaxy of Abell 2597 where Edge et al. (2010) find $L_{[NII]}/L_{FIR}\sim 0.02$. Given that our ALMA observations are missing significant [NII] flux, the true ratio for L2 may be similar to the Abell 2597 value and a factor of 50 to 100 larger than the value seen in most systems (Malhotra et al. 2001, Gracia-Carpio et al. 2011, etc.).

The complex nature of the [NII] and 122 μm continuum emission in SMMJ02399 is arguably the most intriguing new detail revealed by our ALMA data. Previous work argued that the SMMJ02399 may be a single large galaxy (Genzel et al. 2003) with a large rotating molecular gas disk or perhaps a merging system (Ivison et al. 2010, Thomson et al. 2012). Most recently Aguirre et al. (2013) used NIR and optical imaging from HST to model the stellar populations in all of the components of SMMJ02399 (Figure 9). They find that all components contain significant stellar populations. Interestingly, they find L1 and L2SW appear as a single stellar component that they call L1sb. This is distinctly not the case of the dust continuum shown in our ALMA map, which features two emission regions (L2SW and L1) that are separated by ~0.5"—twice the size of the synthesized ALMA beam. The region in our ALMA continuum map between L1 and L2SW, the area where Aguirre et al. (2013) still find significant stellar emission, is undetected down to ~1 mJy. This corresponds to an upper limit on the dust-mass surface density of 23 $M_\odot$ pc$^{-2}$ (assuming $T_d = 45$ K). Aguirre et al. (2013) estimate stellar masses of 6.3 $\times 10^{11} M_\odot$ for the combined L2SW and L1 components, $1.26 \times 10^{10} M_\odot$ for L2 and $5.01 \times 10^9 M_\odot$ for L1N; yet, in our ALMA continuum map L2 and L1N have no detectable dust emission. Even then the dust emission in L2SW is only spatially coincident with about half of the stellar emission of L1sb (see Figure 9). Of course L1 has significant emission from dust, but it is most likely associated with dust heated directly by the central engine. The L1 component is unresolved at ~0.25" scales in our ALMA map, which corresponds to ~ 670 pc in the source plane. While this scale is still consistent with the emission arising in a circumnuclear starburst, but it is likely associated with the AGN and not star formation, since—as mentioned in Section 5.2—the scaled FIR luminosity of L1 matches the expected contribution to the FIR luminosity by the AGN (Lutz et al. 2005). Lastly, if the dust emission in L1 is produced by star formation, then it would be forming stars at a rate that would have made [NII] emission detectable in our ALMA data, yet none is detected.

Clearly, from the differences in the distribution of the stellar and dust emission, SMMJ02399 contains regions with significant dust and stars as well as regions with significant stellar mass but very little dust. All of this suggests that SMMJ02399, while a site of significant obscured star formation, has considerable unobscured star formation as well. The possibility of unobscured star formation is even suggested by early observations by Ivison et al. (1998) who observe a total Hα line flux, uncorrected for dust extinction, of ~$1.4 \times 10^{-17}$ W m$^{-2}$ corresponding to star formation rate of ~2000 M$_\odot$ yr$^{-1}$. This was consistent with the FIR derived SFR given the observations available at the time. Now however, with our ALMA observations, we clearly identify only half of the FIR continuum as being produced in the dusty starburst so that the observed Hα emission over estimates the FIR based SFR by at least a factor of ~3. The Hα SFR can only be explained if there is an unobscured star forming component in addition to the dusty one. Indeed, we have identified that L2 must have significant unobscured star formation since its [NII] emission suggests a SFR of ~ 900 M$_\odot$ yr$^{-1}$ while its dust emission remains undetected.

Arguing that the L2 component has significant star formation traced by strong [NII] emission raises the question as to why the [NII] emission in L2SW is relatively so weak when one typically expects the most intense star formation to occur in dusty environments. First, the star formation rate in L2SW could simply be lower as suggested by comparing the FIR determined SFR in L2SW to the rate in L2 determined via the [NII] line. In this case, the dusty starburst in L2SW may still be more intense (i.e. higher star-formation rate surface-density), since the bulk of the ZEUS-1 [NII] line flux is not detected by ALMA, the star formation producing the line emission must be occurring on large scales and at modest surface densities that can still produce a high star formation rate.

Alternatively, the difference in [NII] line strengths could hint at different starburst ages in the components. In calculating the ionizing flux and subsequent star formation rate from the [NII] line we assume that all of the nitrogen is singly ionized. However, this makes the SFR as traced by [NII] a lower limit. If a star forming region has many more massive and younger stars that are responsible for the ionization, then the ionizing spectrum will extend to energies > 29.6 eV and a significant fraction of the nitrogen will be in higher ionization states. As such, if L2SW represents a more recent burst of star formation, we would expect lower [NII] emission compared to an older starburst. In this paradigm, we predict that the [OIII] 88 μm line, also detected with ZEUS-1 and reported in Ferkinhoff et al. 2010, should come from L2SW, and not L2. Future Band 10 (~ 350 μm) observations with ALMA will allow us to spatially resolve the [OIII] emission and test this prediction.

In light of our discovery that SMMJ02399 contains a significant amount of unobscured star formation in addition to the previously known obscured star formation, we must reexamine our understanding of the nature of SMMJ02399. In other words, we must reexamine if a merger scenario fits all of the observations. To do this we consider the distribution of molecular gas in SMMJ02399. Genzel et al. (2003) image the CO(3-2) emission with a ~5"x 2.5" beam and argue that SMMJ02399 contains a single, large rotating molecular gas disk (M$_{H2}$ = (6.4 ± 2.4) $\times 10^{10}$ M$_\odot$, and M$_{dyn}$ = of $3.2 \times 10^{11}$ M$_\odot$). Ivison et al. (2010) and Thomson et al. (2012) present CO(1-0) observations showing considerable extended emission from molecular gas. They resolve the CO(1-0) emission into a single component, centered on L2SW, that extends over the entire 10 x 10 kpc of the source. The emission traces a total molecular hydrogen mass of $(1.0 \pm 0.3) \times 10^{11}$ M$_\odot$. Both of the CO(1-0) and CO(3-2) observations are consistent with a single molecular gas disk but seem contradictory to the multiple components seen at other wavelengths, especially considering the location of the quasar that is not at the peak of the molecular gas. If however the SMMJ02399 is a major merger nearing final coalescence, like the Antennae galaxies (Arp 244) in the local universe, it can explain all observed properties.

The Antennae are a nearby merging system of two galaxies whose nuclei are ~7 kpc apart. Between the two nuclei, in the overlap region, is a site of significant obscured star formation. The bulk of the molecular gas and dust emission is also co-spatial with the overlap region (Wilson et al. 2000, Schulz et al. 2008). There is also a significant number of young stars outside the overlap region, albeit with slightly older populations (Zhang et al. 2010, Klaas et al. 2010). With the addition of an AGN in one of the galaxy nuclei, this scenario describes SMMJ02399 quite well. In SMMJ02399 the L1 and L2 components are the cores of the merging galaxies separated by ~ 10 kpc in the source plane. L2SW, with the strong dust continuum emission, is the over-lap region and the site of the most recent epoch of star formation. Meanwhile all the stellar emission seen in L2, L1N, and L1sb are the unobscured stellar component formed earlier in the galaxy merger.

In the case of the Cloverleaf the ALMA observations portray a more straightforward picture. The [NII] emission that is detected in the ALMA observations is unresolved in the 0.25" ALMA synthesized beam and spatially associated with the multiply lensed images of the AGN. This suggests that the ALMA obtained line-flux is associated with the NLR of the AGN. The [NII] flux not recovered by ALMA must be extended for it not to be detected. If it is also co-spatial with the Einstein ring in the 122 μm continuum observations, then this missing [NII] flux is associated with the Cloverleaf's starburst. The missing [NII] flux—ZEUS-1 flux minus the ALMA recovered flux—suggest a minimum SFR of 550 ± 50 $M_\odot$ yr$^{-1}$. Considering that this estimate is only a lower limit and that FIR luminosity includes warm dust emission from the AGN, the [NII] derived star formation rate is consistent with the FIR based star formation rate estimates of 950 ± 30 $M_\odot$ yr$^{-1}$. The continuum map of the Cloverleaf is also important, as we resolve the Einstein ring, confirming that there is an important extended component to the source. The high SNR of the ALMA continuum observations enables future work to make a detailed model of the gravitational lens and recover the source image.

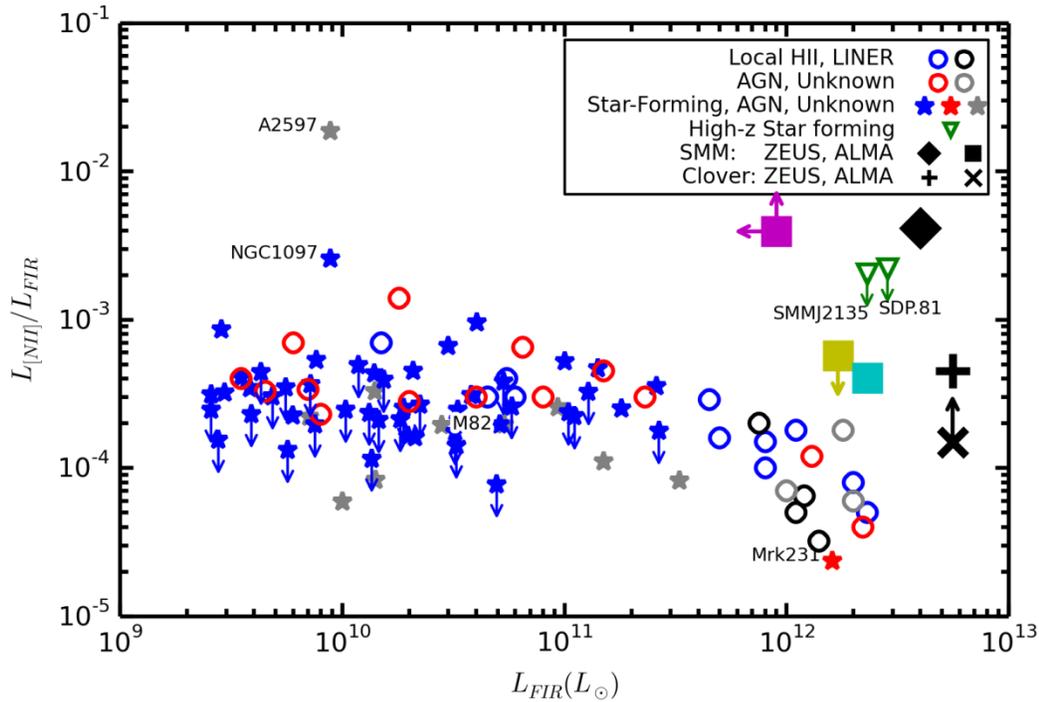

Figure 10: The [NII] 122 μm line-luminosity to FIR luminosity ratio versus the FIR luminosity for SMMJ02399 and the Cloverleaf as well as sources from the literature. The ZEUS derived values (Ferkinhoff et al. 2011) are indicated with a diamond and plus respectively for SMMJ02399 and the Cloverleaf while our new ALMA results are shown with squares and cross. The ZEUS-1 and ALMA values are corrected for lensing. We also include estimates for the SMMJ02399 components: L2 lower-limit (magenta), L1 upper-limit (yellow), and L2SW (cyan). Local sources from Gracia-Carpio et al. (2011) are indicated with open circles, while local sources from Fischer et al. 2010, Edge et al. 2010, Beirao et al. 2010, Vasta et al. 2010, and Malhotra et al 2001 are noted with filled stars. All are colored according to source type. High-z systems reported in Ivison et al. 2010b and Valtchanov et al. 2011 are shown with open triangles.

## 7. Summary

Using ALMA in Cycle-0 of early science observations we observed two high-z systems in their rest frame 122 μm continuum and [NII] 122 μm line emission. Using these ALMA Band 9 (~450 μm, 645 GHz) observations we:

1) Map the dust continuum in both sources at ~0.25" spatial scales, resolving an Einstein ring in the Cloverleaf and two clear components (the quasar and star forming region) in SMMJ02399;
2) Detect the [NII] line in both sources. However, only ~1/5 of the ZEUS-1 line-flux is recovered. The ALMA observations were carried out with a significantly longer baselines than expected and we argue that the missing line flux was resolved out;
3) Confirm the [NII] line flux obtained with ZEUS-1 observations of SMMJ02399 by comparing it to estimates based on the source's free-free emission. Combining its free-free continuum emission, [NII] line flux and [OIII] line flux we determine the N/H abundance in SMMJ02399 is consistent with Milky Way abundances;
4) Determine, even with the small fraction of the [NII] line flux recovered by ALMA, that >80% the ZEUS-1 detected [NII] emission in the Cloverleaf and SMMJ02399 is due to star formation;
5) Confirm that both sources have significant gas ionized by star formation and using the [NII] line flux as a star formation rate indicator, we determine that SMMJ02399 has a significant amount of unobscured star formation. This is the first time the unobscured star formation has been probed in this source;
6) Conclude that SMMJ02399 is a major merger nearing coalescence. Based on our ALMA observations, combined with studies in the literature on the stellar and molecular gas content, SMMJ02399 may represent a high-z analog to the Antennae galaxies.

This work demonstrates the unique power of ALMA for understanding early galaxies. Given the complex nature of the [NII] emission in our sources, resolved studies with ALMA are crucial for understanding single-dish observations of the FIR fine structure lines in high-z galaxies. Furthermore, our study of SMMJ02399 suggests that unobscured star formation in submillimeter galaxies may still be important, regardless of their extreme infrared luminosities. As ALMA begins resolving the dust continuum emission in many high-z systems, we will be able to better assess the relative importance of obscured and unobscured star formation in early galaxies. Detection of the FIR fine structure lines, like the [NII] line, with ALMA will be important to these studies. This work also serves a prime example of the need to approach resolved, interferometric studies cautiously. Given small beams that ALMA can provide, especially at short wavelengths, it is easy to resolve-out significant amounts of a sources flux even at high-z.

This paper makes use of the following ALMA data: ADS/JAO.ALMA# 2011.0.00747.S. ALMA is a partnership of ESO (representing it member states), NSF (USA) and NINS (Japan), together with NRC (CANADA) and an NSC and ASIAA (Taiwan), in cooperation with the Republic of Chile. The Joint ALMA Observatory is operated by ESO, AUI/NRAO and NAOJ. The National Radio Astronomy Observatory is a facility of the National Science Foundation operated under cooperative agreement by Associated Universities, Inc. This research also made use of Astropy, a community-developed core Python package for Astronomy (Astropy Collaboration, 2013).We thank the anonymous referee for help feedback. We also would like to thank the staff at the NRAO and North American ALMA Science Center for their support of these observations and their reduction. Lastly, C. Ferkinhoff thanks R. Ivison for providing the continuum data from Ivison et al. 2010 and to F. Walter for helpful comments on the draft of this paper.